\pdfoutput=1
\documentclass[11pt,a4paper]{article}
\usepackage{moriond,epsfig}

\usepackage{wrapfig,rotating}

\def\be{\begin{equation}}
\def\ee{\end{equation}}
\def\bea{\begin{eqnarray}}
\def\eea{\end{eqnarray}}

\begin{document}
\vspace*{4cm}
\title{DATA PRESERVATION AND LONG TERM ANALYSIS IN HIGH ENERGY PHYSICS}

\author{DAVID M. SOUTH}

\address{Technische Universit\"at Dortmund, Experimentelle Physik V,\\
44221 Dortmund, Germany}

\maketitle\abstracts{
High energy physics data is a long term investment and contains the potential
for physics results beyond the lifetime of a collaboration.
Many existing experiments are concluding their physics programs, and looking
at ways to preserve their data heritage.
Preservation of high-energy physics data and data analysis structures
is a challenge, and past experience has shown it can be difficult if adequate
planning and resources are not provided.
A study group has been formed to provide guidelines for such data preservation
efforts in the future.
Key areas to be investigated were identified at a workshop at DESY
in January 2009, to be followed by a workshop at SLAC in May 2009.
More information can be found at {\tt http://dphep.org}.
}

\section{The Case for Preserving HEP Data}
\label{sec:intro}

A generation of high energy physics (HEP) experiments are concluding
their data taking and winding up their physics programmes.
These include H1 and ZEUS at the $ep$ collider HERA (data taking ended
July 2007), BaBar at the $e^{+}e^{-}$ collider at SLAC (ended April 2008)
and the Tevatron $p\bar{p}$ experiments D{\O} and CDF, who are expected
to stop data taking in 2011.
The experimental data from these experiments still has much to tell us from
the ongoing analyses that remain to be completed, but it may also contain
things we do not yet know about.

%%%

In making the case for data presevation, several scenarios come to mind
where this action would be of benefit to the particle physics community:
We may want to re-do previous measurements with increased precision.
Reduced systematic errors may be possible via new and improved
theoretical calculations (MC models) or newly developed analysis techniques.
We may want to perform new measurements at energies and processes
where no other data are available (or will become available in the future).
If new phenomena are found at the LHC or some other future collider, it may
be useful or even mandatory to go back, if possible, and verify such results
using older data.
To give some specific examples of real data:
The $p\bar{p}$ collisions from Tevatron will provide a contingency for LHC data,
as well as a lower energy point.
The $e^{+}e^{-}$ data from the BaBar and Belle collaborations may still provide
future interest, for comparison to the coming Super-B factory, in particular if
data sets can be combined.
And the $ep$ collisions recorded at HERA represent a unique data set, which is
unlikely to be superseded anytime soon, even considering such future projects
as the LHeC.

%%%

In summary, it may be highly desirable for a collaboration to employ a program
of data preservation, safeguarding the data heritage of the experiment for
continued and future use.
However, implementing such a program is particularly challenging,
as discussed in the next section.

\section{The Challenge of Preserving HEP Data}
\label{sec:challenges}

High energy physics experiments have little or no tradition or clear model
of long term preservation of their data in a meaningful and useful way.
It is in fact likely that most older HEP experiments have simply lost the data.
The preservation of and supported long term access to the data are 
generally not part of the planning, software design or budget of such
experiments.
Additionally, the distribution of the data complicates the task,
with potential headaches arising from ageing hardware where the
data themselves are stored, as well as from unmaintained and
outdated software, which tends to be under the control of the 
(defunct) experiments rather than the associated HEP computing centres.

%%%

Why is this the case?
One of the main assumptions concerning HEP experimental data is often that the
older data will always be superseded by that from the next experiment,
but this is not always the case.
Another sometimes wrong assumption is that the physics potential is
exhausted at the end of the experimental program.
A recent example contradicting such assumptions is the re-analysis of the
JADE experimental $e^{+}e^{-}$ data, using refined theoretical input
and a better simulation, which has lead to a significant improvement in the
determination of the strong coupling, in an energy range that is
still unique \cite{jade}.
A picture of a simulated event in the JADE detector made using the resurrected
event display program, notably now in full colour, is displayed in figure 1.
The recovery and further analysis of the JADE data was however not a
planned endeavour, but rather a private and often arduous but adventurous
initiative, pushed by knowledgeable former members of the collaboration.

\begin{wrapfigure}{r}{70mm}
\begin{center}
  \includegraphics[width=0.4\textwidth]{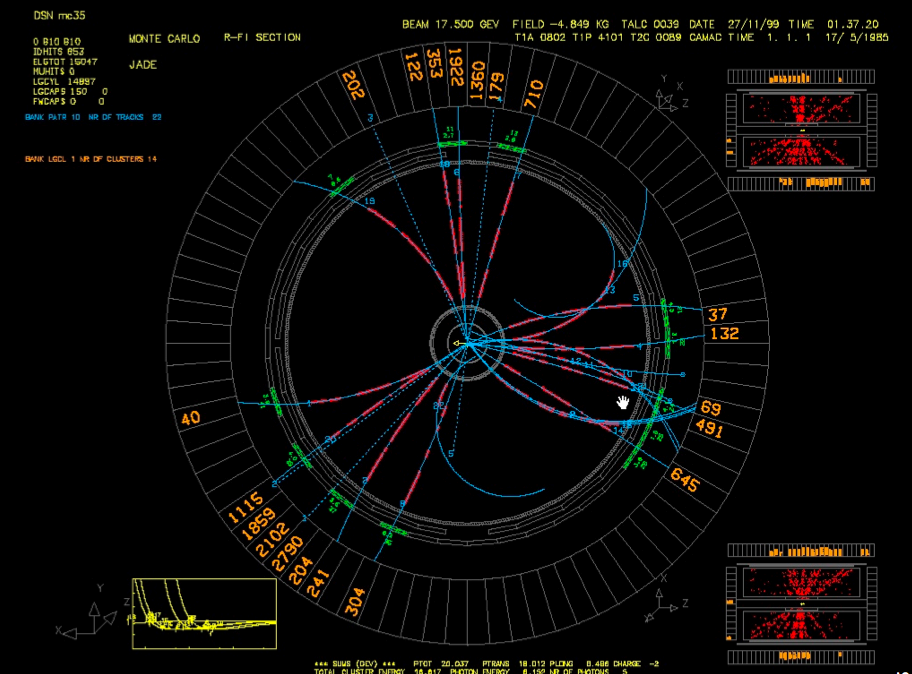}
\end{center}
 \caption{A simulated event in the JADE detector, generated using a refined
	Monte Carlo and reconstructed using revitalised software more than
	ten years after the end of the experiment.}
\end{wrapfigure}

The general status of the LEP data, recorded as recently as the year 2000, is
already a concern, where preservation efforts have run in a rather uncoordinated
and incomplete fashion.
The recovery of all of the required information concerning the LEP data may
become impossible on a time scale of a few years if no dedicated effort is
allocated to the preservation task.
Past experiences like those described above indicate that the definition
of the data should include all the necessary ingredients to retrieve and understand
it in order perform new analyses and that a complete re-analysis is only possible
when all the ingredients can be accounted for.
Furthemore, an early preparation is needed and sufficient resources should be
provided in order to maintain the capability to re-investigate such older data samples.

%%%

Finally, there is the challenge of providing useful open access of HEP data beyond
the walls of the original collaboration.
This is clearly a difficult prospect, with many issues like control, correctness
and reputation of the experiment, not to mention a lack of portability and
the typical state of the documentation within the collaboration.
The implications of such schemes must also be considered in any data 
preservation programme.

\section{A Systematic Approach to Data Preservation in HEP}
\label{sec:dphep}

In order to address this issue in a systematic way, a Study Group on Data
Preservation and Long Term Analysis in HEP was formed at the end of 2008, 
with the aim to clarify the objectives and the means of data persistency in HEP.
The major experiments at colliders: H1, ZEUS, CDF, D¯, BaBar, Belle, Cleo and
BES-III and the associated computing centres at DESY (Germany), Fermilab (USA),
SLAC (USA), KEK (Japan) and IHEP (China) are all represented in the Study Group.
The first workshop of the Study Group took place at DESY in January 2009
some of the key points from this first workshop are summarised below.

\begin{figure}
\begin{center}
 \includegraphics[width=0.95\textwidth]{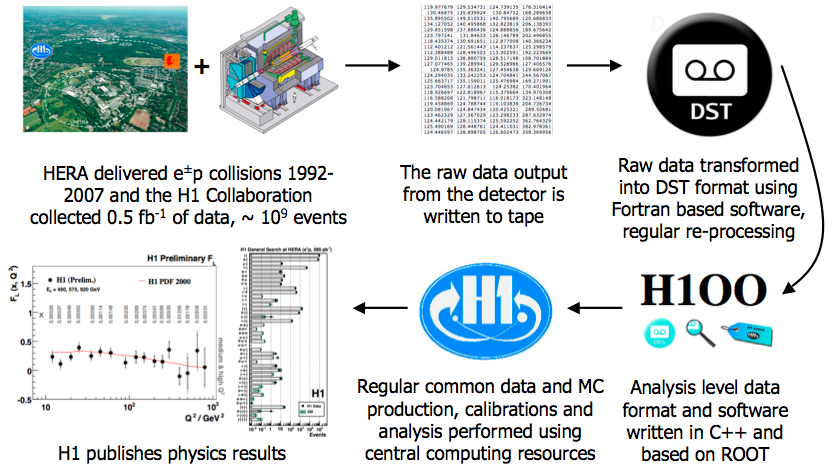}
\end{center}
 \caption{An example of a data analysis model, from the H1 experiment at HERA.
	Similar schemes exist at the other major HEP experiments, as presented
	at the First Data Preservation Workshop at DESY \protect\cite{dplta1}.}
\end{figure}

A comparison of the computing and data analysis models of the experiments in the Study Group
has been performed, including the applicability and adaptability to long term analysis.
Not surprisingly, the models are similar, reflecting the nature of colliding
experiments: an example is given in figure 2, which shows the analysis model
of the H1 experiment at HERA.
Experimental data are organised by events, with increasing abstraction
levels from RAW detector level quantities to ntuple-like data for
physics analysis, and are supported by large simulated Monte Carlo samples.
The software is also organised in a similar manner, with a more conservative part for
reconstruction, reflecting the hardware complexity and a more dynamic part closer
to the analysis level.
The prospective amount of data to be preserved for analysis varies between 0.5 and 10 Pb
per experiment, not huge by today's standards, but nonetheless significant.
The degree of preparation for long term data preservation is diverse among the
experiments, but it is obvious that no preparation was foreseen at an early stage
of the any of the experimental programmes and that any conservation initiative will be
in parallel with the end of analysis at the current experiments.

%%%

It is widely accepted that digital data is a fragile object, when considered in
a long term perspective.
It is also well established that the storage technology should not pose problems
with respect to the amount of data in discussion, and that the main issue will
in fact be the communication between the experiments and the computing centres 
after the end of analyses and/or the collaborations: exactly where roles have not
been clearly defined in the past.
The current preservation model, where the data are simply saved on tapes, runs the
risk that the data may disappear into cupboards while the readout hardware may be
lost or become impractical.
The conservation of tapes is not equivalent to data preservation.
It appears mandatory to define a clear protocol for data preservation,
the items of which should be transparent enough to ensure that the digital
content of an experiment (data and software) is at least accessible.

%%%

The longevity of the analysis software is also an important issue.
The most popular analysis framework is ROOT \cite{root}, which offers many possibilities to
store and document HEP data and has the advantage of a large user community and
a long term commitment of support.
One particular example of software dependence is the use of inherited libraries
(for example CERNLIB or GEANT3) and of commercial software and/or packages that are 
no longer officially maintained, but remain crucial to most of the running experiments.
It would seem advantageous if such vulnerabilities were to be removed from the
software model of the experiments as a first step towards long term stability of
the analysis framework.
Modern techniques of software emulation like virtualisation may also offer promising
features, and exploring such solutions should be a part of the future investigations.

%%%

An increasing awareness of the funding agencies towards the preservation of the
scientific data is noticeable.
In particular, the UE/FP7 funded project PARSE/Insight \cite{parse} recently
conducted a survey of the HEP community, showing that the vast majority of
scientists strongly support the preservation of HEP data.
The next generation of publications database, INSPIRE, offers an attractive option
for extended data storage capabilities, which could be used immediately to enhance
public or private information ranging from scientific articles up to and potentially
including even analysis software and data.

\section{Summary and Future Working Directions}
\label{sec:conclusions}

HEP data is a long term investment and contains a true potential for physics results
beyond the collaborations lifetime.
A study group has been formed to reflect on data preservation and long term analysis
in HEP, more details can be found at {\tt http://dphep.org}.
High energy physics experiments can take some concrete action now and propose models
for data preservation.
The whole process must be supervised by well defined structures and steered by clear
specifications, endorsed by the major laboratories and computing centres.
Technological aspects play an important role, since one of the crucial factors
may indeed be the evolution of the hardware.
A second workshop of the Study Group takes place at SLAC in May 2009, with the aim
of producing a report containing a set of guidelines for further reference on data
preservation and long term analysis in HEP \cite{dplta2}.
%
%Four key areas identfied at the first workshop form the working directions for the
%second, namely: Physics Cases for Data Preservation; Preservation Models; Collaborations,
%Governance and Data Access Policies; Technologies and Facilities.

\end{document}